\documentclass[cameraready]{Interspeech}

\title{Contrastive Regularization for Accent-Robust ASR}

\author[affiliation={1}, orcid=0000-0001-8128-1120]{Van-Phat}{Thai}
\author[affiliation={1}, orcid=0009-0003-5094-3787]{Aradhya}{Dhruv}
\author[affiliation={1,2}, orcid=0000-0001-5156-8171]{Duc-Thinh}{Pham}
\author[affiliation={1}, orcid=0000-0002-7379-8223]{Sameer}{Alam}

\address{
    $^1$ Air Traffic Management Research Institute, Nanyang Technological University, Singapore \\
    $^2$ Center of AI Research, VinUniversity, Vietnam
}

\email{vanphat.thai@ntu.edu.sg,aradhya.dhruv@ntu.edu.sg,dtpham@ntu.edu.sg,sameeralam@ntu.edu.sg}


\usepackage{comment}
\usepackage{amsmath}
\usepackage{multirow}

\begin{document}

\maketitle

\begin{abstract}
ASR systems based on self-supervised acoustic pretraining and CTC fine-tuning achieve strong performance on native speech but remain sensitive to accent variability. We investigate supervised contrastive learning (SupCon) as a lightweight, accent-invariant auxiliary objective for CTC fine-tuning. An utterance-level contrastive loss regularizes encoder representations without architectural modification or explicit accent supervision. Experiments on the L2-ARCTIC benchmark show consistent WER reductions across multiple pretrained encoders, with up to 25 -- 29\% relative reduction under unseen-accent evaluation. Analysis using within-transcript cosine dispersion indicates that SupCon promotes more compact and stable representation geometry under accent variability. Overall, SupCon provides an effective and model-agnostic regularization strategy for improving accent robustness.
\renewcommand{\thefootnote}{}
\footnotetext{The source code and models are publicly available at \url{https://github.com/thaivanphat95/robust-atc-asr}}
\renewcommand{\thefootnote}{\arabic{footnote}}
\end{abstract}
\section{Introduction}
Modern ASR systems based on self-supervised acoustic pretraining and CTC fine-tuning achieve strong performance on benchmarks dominated by native speech~\cite{20w2v2,22wavlm,23whisper}. However, performance degrades substantially for non-native speech, particularly in low-resource or globally deployed settings, due to systematic pronunciation variability that deviates from native speech norms~\cite{18Arctic}. These accent-related variations pose persistent challenges for robust ASR.

To address accent-related variability, ASR research has increasingly focused on the multi-accent setting, where a single model must generalize across diverse accent conditions, many of which may be underrepresented or unseen during training. Existing approaches can be broadly grouped into two categories. Accent-specific methods incorporate accent information explicitly, for example through auxiliary accent classifiers, accent embeddings, or accent-conditioned adaptation mechanisms, enabling specialization to known accents~\cite{19Accent,25Accent-Joint,25Accent-MoE,24Accent-code}. In contrast, accent-invariant methods aim to learn representations that are robust to accent variation without relying on explicit accent supervision, such as modeling accent as a latent factor~\cite{21Accent}, combining supervised ASR with unsupervised learning under low-resource conditions~\cite{25Accent-Unsup}, or augmenting training data with synthetic accented speech~\cite{24Accent-TTS}.

In this work, we investigate supervised contrastive learning (SupCon) as an auxiliary objective for CTC-based ASR fine-tuning. This approach aligns with the accent-invariant direction and does not require architectural modification or explicit accent supervision.
SupCon leverages label supervision to define contrastive pairs, encouraging samples with the same label to form compact clusters while maintaining inter-class separation~\cite{20Supcon}. In natural language processing, this objective has been shown to improve robustness and generalization when applied during fine-tuning of pretrained models~\cite{21Supcon-LLM}.

In speech processing, SupCon has been explored within ASR in a limited number of settings. For accented ASR, Han et al.~\cite{21SupCon} apply SupCon using augmentation-based positive pairs at the character level, while SCaLa~\cite{22Supcon} introduces a phoneme-level contrastive objective relying on forced alignment. Although these approaches demonstrate measurable gains, they typically depend on task-specific pair construction or additional supervision, leaving open how to integrate contrastive regularization more generally into SSL-based ASR fine-tuning. Contrastive objectives have also appeared in neighboring tasks such as speaker verification~\cite{25SupCon}, but these settings differ fundamentally from sequence-to-sequence ASR. As a result, SupCon remains underexplored as a lightweight, model-agnostic auxiliary loss for ASR fine-tuning in multi-accent settings, and existing work does not systematically analyze how contrastive objectives reshape encoder representation geometry or how such changes relate to recognition performance. Drawing inspiration from recent studies on representation geometry, which show that embedding dispersion—measured via average pairwise cosine distance—is strongly correlated with model performance and generalization~\cite{25Dispersion}, we adopt cosine-based dispersion to analyze how SupCon regularizes encoder representations within and across transcripts.

Motivated by these observations, the main contributions of this work are as follows:
\begin{itemize}
    \item We introduce SupCon as an auxiliary loss during CTC fine-tuning, providing a lightweight and model-agnostic regularization strategy for ASR systems based on self-supervised acoustic encoders.
    \item We demonstrate that a standard CTC-based ASR pipeline with SupCon regularization achieves state-of-the-art performance on the L2-ARCTIC benchmark.
    \item We present a geometric analysis of encoder representations using within-transcript cosine dispersion, revealing how contrastive regularization reshapes embedding structure under accent variability.
\end{itemize}

The remainder of the paper is organized as follows. Section~\ref{sec:method} describes the proposed ASR framework and contrastive objective. Section~\ref{sec:exp} details the datasets, evaluation protocols, and experimental settings. Section~\ref{sec:result} presents the main results, ablation studies, and representation analysis. Finally, Section~\ref{sec:con} concludes the paper.

\begin{figure*}
    \centering
    \includegraphics[width=0.9\linewidth]{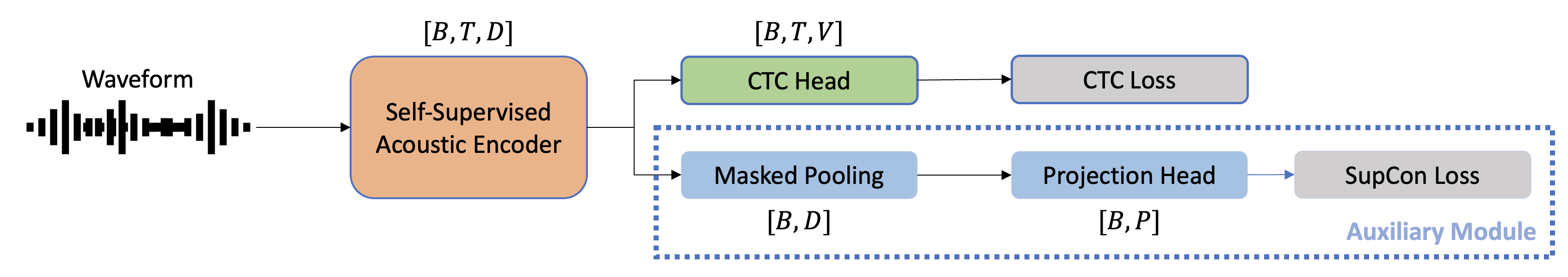}
    \caption{Overview of the proposed training model. A self-supervised acoustic encoder is trained with a primary CTC objective for ASR, while an auxiliary supervised contrastive loss is applied to utterance-level representations obtained via mean pooling of encoder hidden states. The auxiliary module is used only during training and does not affect inference. $B$, $T$, $D$, $V$, and $P$ denote batch size, number of encoder frames, encoder hidden dimension, vocabulary size, and projection dimension, respectively.}
    \label{fig:flow}
\end{figure*}

\section{Methodology}\label{sec:method}
\subsection{Problem Formulation}
As illustrated in Figure~\ref{fig:flow}, let $\mathcal{D} = \{(x_i, y_i)\}_{i=1}^N$ denote a training dataset, where $x_i$ is a raw speech waveform and $y_i$ is the corresponding transcript. Given an utterance $x_i$, a self-supervised pretrained acoustic encoder produces frame-level representations that are shared by the ASR and auxiliary contrastive objectives.

For SupCon, we derive an additional supervision signal $c_i$ from the transcript $y_i$. Specifically, $c_i$ is a transcript-level identifier, such that utterances sharing the same transcript but spoken by different speakers are treated as positive pairs in the contrastive space. This supervision requires no additional annotations beyond standard ASR transcripts.

Our objective is to train an ASR model that maps speech $x_i$ to its transcript $y_i$ using a CTC-based objective, while jointly regularizing the encoder representations with an auxiliary supervised contrastive loss applied at the utterance level. The contrastive objective encourages representations that are robust to speaker and accent variability, whereas inference relies solely on the CTC branch.

\subsection{Acoustic Encoder and CTC Objective}
Following standard SSL-based ASR pipelines, a self-supervised pretrained acoustic encoder processes a batch of $B$ input waveforms. For each utterance $x_i$, the encoder produces a sequence of frame-level representations
\begin{equation}
\mathbf{H}_i = f_{\mathrm{enc}}(x_i) \in \mathbb{R}^{T_i \times D},
\end{equation}
where $T_i$ denotes the number of encoder frames and $D$ is the hidden dimension.

A linear CTC classification head maps the encoder outputs to token-level logits over a vocabulary of size $V$. The model is trained using the standard CTC loss:
\begin{equation}
\mathcal{L}_{\mathrm{CTC}}
=
\frac{1}{B}
\sum_{i=1}^{B}
-\log p_{\mathrm{CTC}}(y_i \mid f_{\mathrm{ctc}}(\mathbf{H}_i)),
\end{equation}
which serves as the primary ASR objective.

\subsection{Supervised Contrastive Loss}
As illustrated in Figure~\ref{fig:flow}, SupCon is applied at the utterance level. Given frame-level encoder outputs $\mathbf{H}_i \in \mathbb{R}^{T \times D}$, we aggregate the valid encoder frames of utterance $i$ into a fixed-dimensional representation using mean pooling:
\begin{equation}
\mathbf{u}_i
=
\frac{1}{\tilde T_i}
\sum_{t=1}^{\tilde T_i}
\mathbf{H}_{i,t}
\in \mathbb{R}^{D},
\end{equation}
where $\tilde T_i$ denotes the number of valid (unpadded) frames. The resulting utterance representation is used as input to the contrastive projection head.

Following standard SupCon practice, each utterance embedding $\mathbf{u}_i$ is mapped to a normalized representation via a lightweight projection head:
\begin{equation}
\mathbf{z}_i = \mathrm{Proj}(\mathbf{u}_i), \qquad
\lVert \mathbf{z}_i \rVert_2 = 1,
\end{equation}
where $\mathrm{Proj}(\cdot)$ denotes a two-layer multilayer perceptron with ReLU activation and $\ell_2$ normalization, and $\mathbf{z}_i \in \mathbb{R}^{P}$.

Given a batch of normalized projections $\{\mathbf{z}_i\}$ and transcript-derived labels $\{c_i\}$, cosine similarity with temperature $\tau$ is defined as
\begin{equation}
s_{ij} = \frac{\mathbf{z}_i^\top \mathbf{z}_j}{\tau}.
\end{equation}

Let $\mathcal{B}=\{1,\dots,B\}$ denote the set of utterance indices in the current batch.
Let $\mathcal{I}\subseteq\mathcal{B}$ denote the set of anchor utterances (e.g., one anchor per transcript).
For each anchor $i \in \mathcal{I}$, we define the comparison set $\mathcal{B}_i = \mathcal{B} \setminus \{i\}$. The positive set is defined as
\begin{equation}
P(i) = \{ j \in \mathcal{B}_i \mid c_j = c_i \}.
\end{equation}

The supervised contrastive loss is defined as
\begin{equation}
\mathcal{L}_{\mathrm{SupCon}}
=
-\frac{1}{|\mathcal{I}|}
\sum_{i \in \mathcal{I}}
\frac{1}{|P(i)|}
\sum_{j \in P(i)}
\log
\frac{\exp(s_{ij})}
{\sum_{k \in \mathcal{B}_i} \exp(s_{ik})}.
\end{equation}

\subsection{Combined Objective with Loss Scheduling}
The final training objective combines the ASR and supervised contrastive losses:
\begin{equation}
\mathcal{L}
=
\mathcal{L}_{\mathrm{CTC}}
+
\lambda_t \mathcal{L}_{\mathrm{SupCon}}.
\end{equation}

To stabilize optimization during early fine-tuning, the contrastive weight is gradually ramped. Let $t$ denote the current training step and $T_{\max}$ the total number of training steps.
The scheduled weight is defined as
\begin{equation}
\lambda_t
=
\lambda \cdot \min\left(1, \frac{t}{r T_{\max}}\right)
\end{equation}
where $\lambda$ is the maximum contrastive weight and $r$ is a fixed ramp ratio.

\begin{table*}[t]
\centering
\caption{Statistics of the L2-ARCTIC dataset under different evaluation settings.
Acc.: accents, Spk.: speakers, Utt.: utterances, Trans.: unique transcripts.}
\label{tab:data_stats}
\begin{tabular}{lcccccccc}
\hline
& \multicolumn{4}{c}{Train} & \multicolumn{4}{c}{Test} \\
\cline{2-5} \cline{6-9}
Setting & \# Acc. & \# Spk. & \# Utt. & \# Trans. &
\# Acc. & \# Spk. & \# Utt. & \# Trans. \\
\hline
\textbf{Unseen-transcript (UT)} &
6 & 18 & $\sim$16k & $\sim$1k &
6 & 6 & $\sim$650 & $\sim$120 \\
\textbf{Unseen-accent (UA)} &
5 & 20 & $\sim$20k & $\sim$1.3k &
1 & 4 & $\sim$4k & $\sim$1.2k \\
\hline
\end{tabular}
\end{table*}

\section{Experimental Setting}\label{sec:exp}
\subsection{Datasets}
We conduct experiments on the L2-ARCTIC~\cite{18Arctic}, a widely used benchmark for non-native and multi-accent ASR. The dataset consists of English speech from non-native speakers across six L1 backgrounds: Arabic, Mandarin, Hindi, Korean, Spanish, and Vietnamese. Each accent group includes four speakers (24 speakers in total), with approximately one hour of speech per speaker. All speakers read the same set of phonetically balanced sentences, enabling controlled evaluation across accents.
Following recent work on multi-accent ASR~\cite{25Accent-MoE}, we adopt two evaluation settings that target complementary generalization scenarios: \emph{unseen-transcript} and \emph{unseen-accent}.

\textbf{Unseen-transcript (UT) evaluation.}  
We follow an 8-fold cross-validation protocol where, for each accent, three speakers are used for training and validation and the remaining speaker is held out for testing. Sentence overlap between splits is avoided, so test utterances contain transcripts not seen during training. This setting evaluates generalization to unseen speakers and unseen linguistic content within known accents.

\textbf{Unseen-accent (UA) evaluation.}  
We adopt a leave-one-accent-out protocol in which five accent groups are used for training and validation, and the remaining accent group is held out entirely for testing. This setting evaluates generalization to previously unseen accents.

We do not consider a combined \emph{unseen-transcript and unseen-accent} setting, as jointly removing transcript and accent overlap complicates interpretation and comparison with prior work. Following existing baselines~\cite{25Accent-MoE}, we therefore evaluate UT and UA separately to isolate linguistic and accent generalization in a controlled and reproducible manner.

\subsection{Training Configuration}

We evaluate wav2vec 2.0~\cite{20w2v2} and WavLM~\cite{22wavlm} encoders in base and large variants using publicly available checkpoints. Experiments are conducted on a single NVIDIA RTX5090 GPU (32GB), with batch sizes of 32 (base) and 16 (large). Training begins with a one-epoch warm-up phase in which only the CTC head is trained (batch size 4) while freezing the encoder to stabilize alignment.
After warm-up, we compare (i) a CTC-only baseline and (ii) joint training with supervised contrastive regularization, using identical architectures and optimization settings. Early stopping is applied based on validation loss with a patience of five epochs.

For SupCon, transcript-balanced batches contain $M$ transcripts with $K$ utterances each ($B = M \times K$). A lightweight projection head with dimension $P = 256$ is used. Optimization follows standard SSL fine-tuning with AdamW (learning rate $1\times10^{-5}$), linear warm-up, and cosine decay. SupCon parameters are set to $\tau = 0.1$, $\lambda = 0.1$, and ramp ratio $r = 0.1$.

During inference, only the CTC branch is used. The projection head is discarded, and decoding employs beam search with a 4-gram language model trained on LibriSpeech clean-360.
\section{Results}\label{sec:result}
\begin{table}[t]
\centering
\caption{Word error rate (WER) on L2-ARCTIC under unseen-transcript (UT) and unseen-accent (UA) evaluation settings.}
\label{tab:main_results}
\begin{tabular}{lcc}
\hline
\multirow{2}{*}{Model} & \multicolumn{2}{c}{WER (\%)} \\
\cline{2-3} & UT & UA \\
\hline
Whisper FT~\cite{25Accent-MoE} & 12.21 & 17.12 \\
MAS-LoRA-QKVO~\cite{25Accent-MoE} & 11.77 & 12.55 \\
\hline
W2V2-Large (CTC) & 10.47 & 9.98 \\
W2V2-Large + SupCon & \textbf{9.14} & \textbf{7.41} \\
\hline
\end{tabular}
\end{table}

\subsection{Main Results}
Table~\ref{tab:main_results} reports word error rate (WER) on the L2-ARCTIC benchmark under unseen-transcript (UT) and unseen-accent (UA) evaluation settings. All results are obtained using identical CTC decoding with a 4-gram language model. The proposed supervised contrastive regularization consistently improves recognition performance over standard CTC fine-tuning.

Using the same wav2vec~2.0 large encoder, SupCon reduces WER from 10.47\% to 9.14\% under the UT setting, corresponding to a relative improvement of 12.7\%. More pronounced gains are observed under the UA setting, where WER is reduced from 9.98\% to 7.41\%, yielding a relative improvement of 25.8\%. This indicates that utterance-level supervised contrastive regularization is particularly effective for improving generalization across accent conditions.

We further observe a clear contrast between accent-aware adaptation and the proposed accent-invariant regularization. Methods based on Whisper fine-tuning and accent-aware adaptation achieve strong performance under the UT setting, where training and test data share accent groups, but degrade noticeably under the UA setting. In contrast, supervised contrastive regularization yields its largest gains under UA evaluation, suggesting that encouraging accent-invariant representation geometry during fine-tuning is especially beneficial for generalization to previously unseen accents. Notably, these improvements are achieved without modifying the underlying ASR architecture or introducing explicit accent supervision.

\subsection{Ablation Study}
Table~\ref{tab:ablation} presents an ablation study analyzing the effects of SupCon and 4-gram LM decoding across four acoustic encoders under UT and UA evaluation settings.
Across all architectures, SupCon consistently reduces WER relative to the CTC-only baseline under both greedy and LM-based decoding. These gains hold for both base and large models, indicating that the proposed auxiliary objective is robust to model scale. Improvements are most pronounced under the UA setting, supporting the hypothesis that utterance-level contrastive regularization promotes accent-invariant representations.

Relative gains from SupCon are generally larger for wav2vec~2.0 than for WavLM. A plausible explanation is that WavLM employs stronger pretraining objectives that encourage context-aware and sequence-level representations, which may already provide implicit regularization during fine-tuning. Consequently, the additional regularization introduced by SupCon yields more modest marginal improvements. A more detailed analysis of these interactions is left for future work.

The impact of external LM decoding varies across models and settings. While LM decoding often improves performance, particularly for larger models and under UA, it may offer limited or slightly negative gains in some UT cases. Importantly, SupCon consistently improves performance regardless of decoding strategy, suggesting that SupCon and LM decoding act as complementary but largely independent factors.

Finally, WavLM-based models generally outperform wav2vec~2.0 under UA but show higher WER under UT. This behavior likely reflects characteristics of the L2-ARCTIC corpus, where speakers across accents read overlapping sentence sets, favoring models with stronger sequence-level modeling. This effect is orthogonal to the gains introduced by supervised contrastive regularization.

\begin{table}[t]
\centering
\caption{Ablation study of SupCon on L2-ARCTIC. Gdy: greedy decoding; LM: 4-gram LM decoding.}
\label{tab:ablation}
\begin{tabular}{llcccc}
\hline
\multirow{2}{*}{Model} & \multirow{2}{*}{Objective}
& \multicolumn{2}{c}{UT} & \multicolumn{2}{c}{UA} \\
\cline{3-6}
& & Gdy & LM & Gdy & LM \\
\hline
\multirow{2}{*}{W2V2-B}
& CTC        & 18.48 & 19.55 & 18.02 & 18.44 \\
& SupCon    & \textbf{17.40} & \textbf{17.49} & \textbf{15.26} & \textbf{14.88} \\
\hline
\multirow{2}{*}{W2V2-L}
& CTC        & 11.58 & 10.47 & 10.91 & 9.98 \\
& SupCon    & \textbf{9.7} & \textbf{9.14} & \textbf{7.74} & \textbf{7.41} \\
\hline
\multirow{2}{*}{WavLM-B}
& CTC        & 25.36 & 18.35 & 15.51 & 11.09 \\
& SupCon    & \textbf{24.68} & \textbf{18.03} & \textbf{13.80} & \textbf{9.83} \\
\hline
\multirow{2}{*}{WavLM-L}
& CTC        & 18.55 & 12.47 & 12.16 & 7.99 \\
& SupCon    & \textbf{18.36} & \textbf{12.30} & \textbf{11.27} & \textbf{6.68} \\
\hline
\end{tabular}
\end{table}

\begin{figure}[t]
    \centering
    \includegraphics[width=\linewidth]{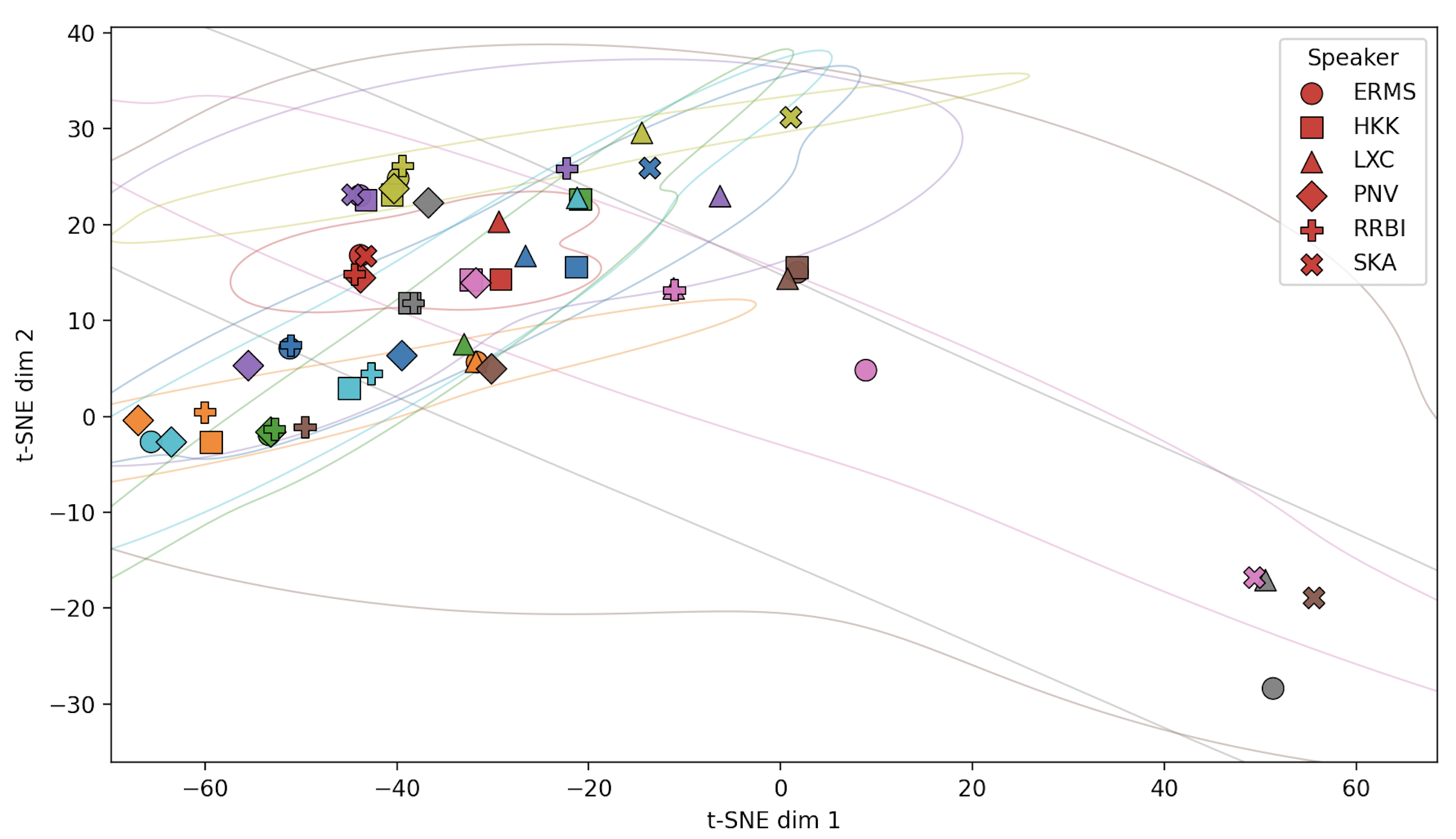}
    \includegraphics[width=\linewidth]{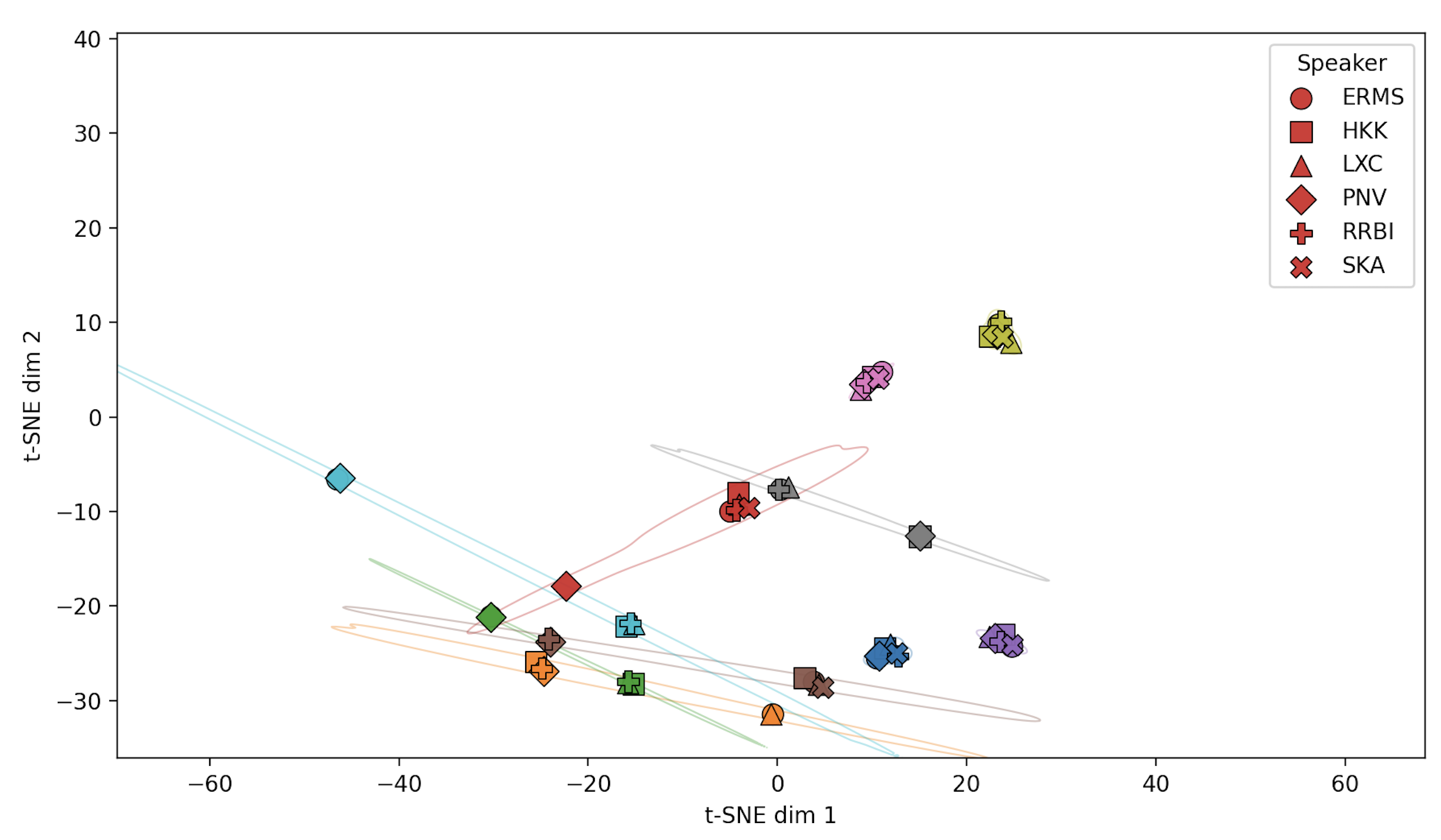}
    \caption{Illustrative t-SNE visualizations of utterance-level encoder embeddings for a shared subset of transcripts. Top: W2V2-Large (CTC). Bottom: W2V2-Large + SupCon. Colors denote transcript identifiers, and marker shapes indicate speakers (each associated with a distinct L1 accent). Transcripts are selected for visualization only, quantitative analysis is reported separately using within-transcript cosine dispersion.}
    \label{fig:tsne_dispersion}
\end{figure}

\subsection{Accent-wise Analysis}\label{subsec:cosine-dispersion}
To assess how consistently the acoustic encoder represents identical linguistic content across accents, we measure the dispersion of utterance-level representations within each transcript. Unlike prior work~\cite{25Dispersion}, which analyzes \emph{global} dispersion across contexts, we focus on \emph{within-transcript} dispersion, where lower values indicate stronger content invariance across speakers and accents.
Let $\mathbf{u}_i \in \mathbb{R}^D$ denote the utterance-level representation obtained by mean pooling the final encoder outputs $\mathbf{H}_i$, prior to the contrastive projection head. Each utterance $i$ is associated with a transcript identifier $c_i$. For a given transcript $c$, we define
\begin{equation}
\mathcal{U}_c = \{ i \mid c_i = c \}, \quad n_c = |\mathcal{U}_c|,
\end{equation}
and consider only transcripts with $n_c \ge 2$.
The within-transcript dispersion is defined as pairwise cosine distance:
\begin{equation}
D(c)
=
\frac{2}{n_c (n_c - 1)}
\sum_{\substack{i,j \in \mathcal{U}_c \\ i < j}}
\left(
1 -
\frac{\mathbf{u}_i^\top \mathbf{u}_j}
{\lVert \mathbf{u}_i \rVert_2 \lVert \mathbf{u}_j \rVert_2}
\right).
\end{equation}
Smaller values of $D(c)$ indicate tighter clustering of utterances with identical transcripts, independent of magnitude.

Figure~\ref{fig:tsne_dispersion} illustrates this effect using t-SNE on a shared subset of transcripts. Compared to W2V2-Large (CTC), the SupCon model produces more compact transcript-level clusters, suggesting improved invariance to speaker and accent variation.
Quantitatively, SupCon consistently reduces within-transcript dispersion across 115 transcripts: the mean decreases from 0.0518 to 0.0430, the median from 0.0460 to 0.0406, and the standard deviation from 0.0254 to 0.0213, corresponding to a 17\% relative reduction in mean dispersion. This reduction aligns with the observed improvements in WER, indicating that SupCon promotes more content-invariant encoder representations without disrupting alignment-sensitive decoding.

L2-ARCTIC contains repeated transcripts spoken by multiple speakers, enabling utterance-level SupCon in this benchmark through transcript-level positive pairs. Importantly, improvements are observed under both UT and UA settings, indicating that the gains are not due to transcript memorization but to improved accent-invariant representation geometry. In scenarios where naturally repeated transcripts are unavailable, positive sets may be constructed through transcript-similarity grouping or synthetic variants, enabling utterance-level contrastive learning beyond benchmark conditions.

\section{Conclusion}\label{sec:con}
This paper demonstrates that supervised contrastive learning is an effective utterance-level regularizer for ASR fine-tuning. Without modifying model architectures or pretraining procedures, the proposed approach improves accent robustness, stabilizes encoder representations, and yields consistent WER reductions across multiple self-supervised acoustic models. Future work will explore alternative positive-pair definitions, including transcript-similarity grouping and synthetic variants for settings where naturally repeated transcripts are unavailable.

\section{Acknowledgments}

\ifcameraready
     This research is supported by the National Research Foundation, Singapore, and the Civil Aviation Authority of Singapore, under the Aviation Transformation Programme. Any opinions, findings and conclusions or recommendations expressed in this material are those of the author(s) and do not reflect the views of National Research Foundation, Singapore and the Civil Aviation Authority of Singapore.
\else
\fi

\section{Generative AI Use Disclosure}
During the preparation of this manuscript, the author(s) used ChatGPT to check grammar, spelling, and syntax errors. After using this tool, the author(s) reviewed and edited the content as needed and take(s) full responsibility for the content of the published article.

\bibliographystyle{IEEEtran}
\bibliography{mybib}

\end{document}